\documentclass[prl,aps,nofootinbib,twocolumn]{revtex4}
\usepackage[dvips]{graphicx}
\usepackage{amssymb}
\usepackage{amsmath}
\usepackage{amsmath, amsthm, amssymb, mathrsfs}
\usepackage{pstricks}

\begin{document}

\title{Does three-tangle properly quantify the three-party entanglement for mixture of  
Greenberger-Horne-Zeilinger states?} 
\author{Eylee Jung, DaeKil Park}

\affiliation{Department of Physics, Kyungnam University, Masan,
631-701, Korea}

\author{Jin-Woo Son}

\affiliation{Department of Mathematics, Kyungnam University, Masan,
631-701, Korea}

\vspace{1.0cm}

\begin{abstract}
Some mixed states composed of only GHZ states can be expressed in terms of only W-states.
This fact implies that such states have vanishing three-tangle. One of such rank-3 states, 
$\Pi_{GHZ}$, is 
explicitly presented in this paper. These results are used to compute analytically
the three-tangle of a rank-4 mixed state $\sigma$ composed of four GHZ states. This analysis
with considering Bloch sphere $S^{16}$ of $d=4$ qudit system allows us to derive the 
hyper-polyhedron. It is shown that the states in this hyper-polyhedron have vanishing 
three-tangle.
Computing the one-tangles for $\Pi_{GHZ}$ and $\sigma$, we prove the 
monogamy inequality explicitly. Making use of the fact that the three-tangle of $\Pi_{GHZ}$
is zero, we try to explain why the W-class in the whole mixed states is not of measure
zero contrary to the case of pure states.
\end{abstract}


\maketitle

Nowadays, it is well-known that entanglement is the most valuable physical resource for
the quantum information processing such as  
quantum teleportation\cite{bennett93}, superdense coding\cite{bennett92}, 
quantum cloning\cite{gisin97-1}, quantum algorithms\cite{shor94}, quantum 
cryptography\cite{fuchs97}, and quantum computer technology\cite{vidal03-1}. Thus, it is 
highly important to understand the various properties of the mutipartite entanglement of the 
quantum states.

The main obstacle for characterizing the entanglement of the multipartite state is 
its calculational difficulties even if original definition of the entanglement measure
itself is comparatively simple. In addition, computation of the entanglement for the 
multipartite mixed states  is much more difficult than that for the pure states, mainly due
to the fact that the entanglement for the mixed states, in general, is defined by a 
convex-roof extension\cite{bennett96}. In order to compute the entanglement explicitly
for the mixed states, therefore, we should find an optimal decomposition of the given
mixed state, which provides a minimum value of the entanglement over all possible 
ensembles of pure states. However, there is no general way for finding the optimal 
decomposition for the arbitrary mixed states except bipartite cases\cite{form2}.
Thus, it becomes a central issue for the computation of the mixed state entanglement.

Few years ago, fortunately, Wootters\cite{form2} has shown how to construct 
the optimal decompositions for the most simple bipartite cases. 
This enables us to be able to compute the concurrence, one of the entanglement measure, 
analytically for the arbitrary $2$-qubit mixed states. It also makes it possible to understand 
more deeply the role of the entanglement in the real quantum information 
processing\cite{08-mixed}. Most importantly, it becomes a basis for the quantification of 
three-party entanglement called residual entanglement or three-tangle\cite{tangle1}.
Thus, it is extremely important to find a calculation tool for the three-tangle if one 
wants to take a step toward a fundamental issue, i.e. characterization of the mutipartite mixed 
state entanglement.

It is well-known\cite{dur00-1} that the three-qubit pure states can be classified by 
product states ($A-B-C$), biseparable states ($A-BC, B-AC, C-AB$) and true tripartite
states ($ABC$) through stochastic local operation and classical communication(SLOCC).
The true tripartite states consist of two different classes, GHZ-class and W-class, where
\begin{equation}
\label{ghzandw}
\begin{array}{l}
|GHZ\rangle = \frac{1}{\sqrt{2}} \left( |000\rangle + |111\rangle \right)            \\
|W\rangle = \frac{1}{\sqrt{3}} \left( |001\rangle + |010\rangle + |100\rangle \right).
\end{array}
\end{equation}
Since the three-tangle $\tau_{3}$ for the pure state 
$|\psi\rangle = \sum_{i,j,k=0}^1 a_{ijk} |ijk\rangle$ is defined as\cite{tangle1} 
\begin{equation}
\label{3-tangle-1}
\tau_{3} = 4 |d_1 - 2 d_2 + 4 d_3|
\end{equation}
with
\begin{equation}
\label{3-tangle-2}
\begin{array}{lll}
& &d_1 = a^2_{000} a^2_{111} + a^2_{001} a^2_{110} + a^2_{010} a^2_{101} +
                                                              a^2_{100} a^2_{011}
                                                              \\
& &d_2 = a_{000} a_{111} a_{011} a_{100} + a_{000} a_{111} a_{101} a_{010}
                                                              \\
& &\qquad 
       + a_{000} a_{111} a_{110} a_{001} + a_{011} a_{100} a_{101} a_{010}
                                                              \\
& &\qquad 
       + a_{011} a_{100} a_{110} a_{001} + a_{101} a_{010} a_{110} a_{001}
                                                              \\
& &d_3 = a_{000} a_{110} a_{101} a_{011} + a_{111} a_{001} a_{010} a_{100},
\end{array}
\end{equation}
it is easy to show that the product and biseparable states have zero three-tangle.
This fact implies that the three-tangle is a genuine measure for the three-party entanglement.

However, there is a crucial defect in the three-tangle as a three-party entanglement measure.
While the three-tangle for the GHZ state is maximal, i.e. $\tau_{3} (GHZ) = 1$, it vanishes for
the W-state. This means that the three-tangle does not properly quantify the three-party
entanglement for the W-type states. The purpose of this letter is to show that besides W-type
states the three-tangle $\tau_3$ does not properly quantify the three-party entanglement for a
rank-$3$ mixtures composed of only three GHZ-type states. 

Recently, the 
three-tangle for rank-$2$ mixture of GHZ and W states is analytically computed\cite{tangle2}.
In Ref.\cite{tangle3}, furthermore, the three-tangle for the rank-$3$ mixture of GHZ, W, and
inverted W states is also analytically computed. In this letter we start with showing that a 
mixed state
\begin{equation}
\label{tangle-x-1}
\begin{array}{l}
\Pi_{GHZ} = \frac{1}{3} \big[    \;
                                  \left| GHZ,2+\rangle \langle GHZ,2+ \right|
                                                                              \\
                            \quad
                                + \left| GHZ,3+\rangle \langle GHZ,3+ \right|
                                + \left| GHZ,4+\rangle \langle GHZ,4+ \right|
                                  \;
                        \big]
\end{array}
\end{equation}
has vanishing three-tangle, where we define for later use as following:
\begin{equation}
\label{summary2}
   \begin{array}{l}
|GHZ, 1\pm\rangle = \frac{1}{\sqrt{2}} \left( |000\rangle \pm |111\rangle \right)
                                                                                    \\
|GHZ, 2\pm\rangle = \frac{1}{\sqrt{2}} \left( |001\rangle \pm |110\rangle \right)
                                                                                    \\
|GHZ, 3\pm\rangle = \frac{1}{\sqrt{2}} \left( |010\rangle \pm |101\rangle \right) \hspace{1.0cm}
                                                                                    \\
|GHZ, 4\pm\rangle = \frac{1}{\sqrt{2}} \left( |011\rangle \pm |100\rangle \right). \hspace{1.0cm}
   \end{array}
\end{equation}

Let us consider a pure state
\begin{equation}
\label{appen-1}
   \begin{array}{l}
      |J(\theta_1, \theta_2)\rangle
         =
             \frac{1}{\sqrt{3}} |GHZ,2+\rangle
           - \frac{1}{\sqrt{3}} e^{i \theta_1} |GHZ,3+\rangle
                                                                   \\
         \qquad\qquad\quad
           - \frac{1}{\sqrt{3}} e^{i \theta_2} |GHZ,4+\rangle.
   \end{array}
\end{equation}
Then, it is easy to show that the three-tangle of $|J(\theta_1, \theta_2)\rangle$ is 
\begin{equation}
\label{appen-2}
\tau_3(\theta_1, \theta_2) = \frac{1}{9} |1 - \left(e^{i \theta_1} - e^{i \theta_2} \right)^2 |
                     |1 - \left(e^{i \theta_1} + e^{i \theta_2} \right)^2 |,
\end{equation}
which vanishes when
\begin{equation}
\label{appen-3}
   \left( \theta_1, \theta_2 \right)
      =
         \left\{
            \begin{array}{l}
               \displaystyle
                  \left( \phantom{2}\pi/3, 2\pi/3 \right),
                  \left( 5\pi/3, 4\pi/3 \right)  
                                                                   \\
               \displaystyle
                  \left( 2\pi/3, \phantom{2}\pi/3 \right),
                  \left( 4\pi/3, 5\pi/3 \right)  
                                                                   \\
               \displaystyle
                  \left( \phantom{2} \pi/3, 5\pi/3 \right),
                  \left( 5\pi/3, \phantom{2}\pi/3 \right)  
                                                                   \\
               \displaystyle
                  \left( 2\pi/3, 4\pi/3 \right),
                  \left( 4\pi/3, 2\pi/3 \right)  
            \end{array}
         \right\}.
\end{equation} 
Moreover, one can show straightforwardly that $\Pi_{GHZ}$ can be decomposed into
\begin{equation}
\label{appen-4}
\begin{array}{l}
   \Pi_{GHZ}
      =
         \frac{1}{8}
         \big[ \;
               \left|\right. J\left( \pi/3, 2\pi/3 \right) \rangle
               \langle J\left( \pi/3, 2\pi/3 \right) \left|\right.
                                                                        \\
             \qquad\qquad\quad
             + \left|\right. J\left(5\pi/3, 4\pi/3 \right) \rangle
               \langle J\left(5\pi/3, 4\pi/3 \right) \left|\right. 
                                                                        \\
             \qquad\qquad\quad
             + \left|\right. J\left( \pi/3, 5\pi/3 \right) \rangle
               \langle J\left( \pi/3, 5\pi/3 \right) \left|\right.
                                                                        \\
             \qquad\qquad\quad
             + \left|\right. J\left(2\pi/3, 4\pi/3 \right) \rangle
               \langle J\left(2\pi/3, 4\pi/3 \right) \left|\right. 
                                                                        \\
             \qquad\qquad\quad
             + \;\textrm{terms with exchanged arguments} \;
         \big]. 
\end{array}
\end{equation}
Combining Eq.(\ref{appen-3}) and (\ref{appen-4}), one can show that Eq.(\ref{appen-4}) is 
the optimal decomposition of $\Pi_{GHZ}$ and its three-tangle is zero:  
\begin{equation}
\label{appen-5}
\tau_3 \left(\Pi_{GHZ}\right) = 0.
\end{equation}
The reason why $\Pi_{GHZ}$ has vanishing three-tangle is that the optimal ensembles given
in Eq. (\ref{appen-4}) are all W-states. Therefore, $\Pi_{GHZ}$ can also be expressed 
in terms of only W-states.
As a result, we encounter a very strange situation that $\Pi_{GHZ}$ has vanishing three- and 
two-tangles\footnote{It is easy to show that ${\cal C}^2_{AB}$ and ${\cal C}^2_{AC}$
are zero, where ${\cal C}$ is concurrence for corresponding reduced states.}, but 
non-vanishing one-tangle
\begin{equation}
\label{one-tangle-1}
4 \min \left[\mbox{det} \left( \mbox{Tr}_{BC} \Pi_{GHZ} \right) \right] = \frac{5}{9}.
\end{equation} 

For comparison one can compute $\pi$-tangle\cite{ou07-1}, another three-party entanglement
measure defined in terms of the global negativities\cite{vidal01-1}. It is easy to show that the
$\pi$-tangle of $\Pi_{GHZ}$ is not vanishing but $1/9$. This fact seems to indicate that the 
three-tangle does not properly reflect the three-party entanglement for GHZ-type states as 
well as W-type states.

We can use Eq.(\ref{appen-5}) for computing the three-tangles of the higher-rank mixed states.
For example, let us consider the following rank-$4$ state
\begin{equation}
\label{rank-4-1}
\sigma = x |GHZ,1+\rangle \langle GHZ,1+| + (1 - x) \Pi_{GHZ}
\end{equation}
with $0 \leq x \leq 1$.
In order to compute the three-tangles for $\sigma$ we first consider a pure state
\begin{equation}
\label{tangle-x-2}
   \begin{array}{l}
      |X(x, \varphi_1, \varphi_2, \varphi_3)\rangle
         =
            \sqrt{x} |GHZ,1+\rangle
                                                                                  \\
          - \sqrt{\frac{1-x}{3}} 
            \left(
               \begin{array}{l}
                   \phantom{+}
                     e^{i \varphi_1} |GHZ,2+\rangle 
                   + e^{i \varphi_2} |GHZ,3+\rangle
                                                                                  \\
                   + e^{i \varphi_3} |GHZ,4+\rangle
               \end{array}
            \right).
   \end{array}
\end{equation}
Then it is easy to show that the three-tangle of 
$|X(x, \varphi_1, \varphi_2, \varphi_3)\rangle$ becomes
\begin{equation}
\label{tangle-x-3}
\begin{array}{l}
   \tau_3 \left(|X(x, \varphi_1, \varphi_2, \varphi_3)\rangle \right)
                                                  \\
      \quad
      =
         \left|
            \begin{array}{l}
               x^2
             + \frac{(1-x)^2}{9}
               \left(
                     e^{4i\varphi_1}
                   + e^{4i\varphi_2}
                   + e^{4i\varphi_3}
               \right)
                                                  \\
             - \frac{2}{3} x (1-x)
               \left(
                     e^{2i\varphi_1}
                   + e^{2i\varphi_2}
                   + e^{2i\varphi_3}
               \right)
                                                  \\
             - \frac{2}{9} (1-x)^2
               \left(
                  \begin{array}{l}
                        e^{2i (\varphi_1 + \varphi_2)}
                      + e^{2i (\varphi_1 + \varphi_3)}
                                                       \\
                      + e^{2i (\varphi_2 + \varphi_3)}
                  \end{array}
               \right)
                                                  \\
             - \frac{8 \sqrt{3}}{9} \sqrt{x (1 - x)^3}
               e^{i (\varphi_1 + \varphi_2 + \varphi_3)}
            \end{array}
         \right|.
\end{array}
\end{equation}

The vectors $ |X(x, \varphi_1, \varphi_2, \varphi_3)\rangle$ has following properties. 
The three-tangle of it has the largest zero at $x=x_0 \equiv 3/4$ and 
$\varphi_1 = \varphi_2 = \varphi_3 = 0$. The vectors $|X(x, 0, 0, 0)\rangle$,
$|X(x, 0, \pi, \pi)\rangle$, $|X(x, \pi, 0, \pi)\rangle$ and $|X(x, \pi, \pi, 0)\rangle$ have
same three-tangles. Finally, $\sigma$ can be decomposed into
\begin{equation}
\label{tangle-x-4}
\begin{array}{l}
   \sigma
      =
         \frac{1}{4}
         \big[
            \;
                  |X(x, 0, 0, 0)\rangle \langle X(x, 0, 0, 0)|
                                                                       \\
                + |X(x, 0, \pi, \pi)\rangle \langle X(x, 0, \pi, \pi)|
                + |X(x, \pi, 0, \pi)\rangle \langle X(x, \pi, 0, \pi)|
                                                                        \\
                + |X(x, \pi, \pi, 0)\rangle \langle X(x, \pi, \pi, 0)|
            \;
         \big].
\end{array}
\end{equation}
When $x \leq x_0$, one can construct the optimal decomposition in the following form:
\begin{equation}
\label{tangle-x-5}
   \begin{array}{l}
      \sigma
         =
            \frac{x}{4 x_0}
            \big[
               \;
                  |X(x_0, 0, 0, 0)\rangle \langle X(x_0, 0, 0, 0)|
                                                                           \\
               \qquad
                + |X(x_0, 0, \pi, \pi)\rangle \langle X(x_0, 0, \pi, \pi)|
                                                                           \\
               \qquad
                + |X(x_0, \pi, 0, \pi)\rangle \langle X(x_0, \pi, 0, \pi)|
                                                                           \\
               \qquad
                + |X(x_0, \pi, \pi, 0)\rangle \langle X(x_0, \pi, \pi, 0)|
               \;
            \big]
                                                                           \\
               \qquad
          + \frac{x_0 - x}{x_0} \Pi_{GHZ}.
   \end{array}
\end{equation}
Since $\Pi_{GHZ}$ has the vanishing three-tangle, one can show easily 
\begin{equation}
\label{tangle-x-6}
\tau_3 (\sigma) = 0  \hspace{1.0cm} \mbox{when} \hspace{.2cm} x \leq x_0 = 3/4.
\end{equation}

Now, let us consider the three-tangle of $\sigma$ in the region
$x_0 \leq x \leq 1$. Since Eq.(\ref{tangle-x-4}) is an optimal decomposition at $x=x_0$,
one can conjecture that it is also optimal in the region $x_0 \leq x$. As will be shown
shortly, however, this is not true at the large-$x$ region. If we compute the three-tangle 
under the condition that Eq.(\ref{tangle-x-4}) is optimal at $x_0 \leq x$, its expression
becomes
\begin{equation}
\label{tangle-x-7}
\alpha_I (x) = x^2 - \frac{1}{3} (1 - x)^2 - 2 x (1 - x) - \frac{8 \sqrt{3}}{9}
\sqrt{x (1 - x)^3}.
\end{equation}
However, one can show straightforwardly that $\alpha_I (x)$ is not a convex function
in the region $x \geq x_*$, where
\begin{equation}
\label{tangle-x-8}
x_* = \frac{1}{4} \left( 1 + 2^{1/3} + 4^{1/3} \right) \approx 0.961831.
\end{equation}
Therefore, we need to convexify $\alpha_I (x)$ in the region $x_1 \leq x \leq 1$ to make 
the three-tangle to be convex function, where $x_1$ is some number between 
$x_0$ and $x_*$. The number $x_1$ will be determined shortly.

In the large $x$-region one can derive the optimal decomposition in a form:
\begin{equation}
\label{tangle-x-9}
   \begin{array}{l}
      \sigma    
         =
            \frac{1-x}{4 (1-x_1)}
            \big[
               \;
                  |X(x_1, 0, 0, 0)\rangle \langle X(x_1, 0, 0, 0)|
                                                                            \\
               \qquad\qquad
                + |X(x_1, 0, \pi, \pi)\rangle \langle X(x_1, 0, \pi, \pi)|
                                                                            \\
               \qquad\qquad
                + |X(x_1, \pi, 0, \pi)\rangle \langle X(x_1, \pi, 0, \pi)|
                                                                            \\
               \qquad\qquad
                + |X(x_1, \pi, \pi, 0)\rangle \langle X(x_1, \pi, \pi, 0)|
               \;
            \big]
                                                                            \\
               \qquad\qquad
          + \frac{x - x_1}{1 - x_1}
            |GHZ,1+\rangle \langle GHZ,1+|
   \end{array}
\end{equation}
which gives a three-tangle as 
\begin{equation}
\label{tangle-x-10}
\alpha_{II} (x, x_1) = \frac{1 - x}{1 - x_1} \alpha_I (x_1) + \frac{x - x_1}{1 - x_1}.
\end{equation}
Since $d^2 \alpha_{II} / dx^2 = 0$, there is no convex problem if $\alpha_{II} (x, x_1)$
is a three-tangle in the large-$x$ region.
The constant $x_1$ can be fixed from the condition of minimum $\alpha_{II}$, 
i.e. $\partial \alpha_{II} (x, x_1) / \partial x_1 = 0$, which gives
\begin{equation}
\label{tangle-x-11}
x_1 = \frac{1}{4} (2 + \sqrt{3}) \approx 0.933013.
\end{equation}
As expected, $x_1$ is between $x_0$ and $x_*$. Thus, finally the three-tangle for 
$\sigma$ becomes
\begin{eqnarray}
\label{tangle-x-12}
\tau_3 (\sigma) = \left\{     \begin{array}{cc}
                        0   & \hspace{1.0cm}  x \leq x_0    \\
                        \alpha_I (x)    & \hspace{1.0cm}   x_0 \leq x \leq x_1    \\
                        \alpha_{II} (x, x_1)    & \hspace{1.0cm}    x_1 \leq x \leq 1
                           \end{array}                                    \right.
\end{eqnarray}
and the corresponding optimal decompositions are Eq.(\ref{tangle-x-5}), Eq.(\ref{tangle-x-4})
and Eq.(\ref{tangle-x-9}) respectively. In order to show Eq.(\ref{tangle-x-12}) is genuine 
optimal, first we plot $x$-dependence of Eq.(\ref{tangle-x-3}) for various 
$\varphi_i \hspace{.2cm} (i=1, 2, 3)$. These curves have been referred as the characteristic 
curves\cite{oster07}. Then, one can show, at least numerically, that Eq.(\ref{tangle-x-12}) is 
a convex hull of the minimum of the characteristic curves, which implies that 
Eq.(\ref{tangle-x-12}) is genuine three-tangle for $\sigma$.

It is straightforward to show that the mixture $\sigma$ has vanishing two-tangles,
i.e. ${\cal C}_{AB} = {\cal C}_{AC} = 0$, but non-vanishing one-tangle
\begin{equation}
\label{boso-2}
   {\cal C}^2_{A(BC)} (\sigma)
      = 
         \frac{1}{9} \left( 5 - 4 x + 8 x^2 - 8 \sqrt{3 x (1 - x)^3} \right).
\end{equation}
Thus, the monogamy inequality $\tau_3 + {\cal C}^2_{AB} + {\cal C}^2_{AC} \leq {\cal C}^2_{A(BC)}$
holds for the rank-4 mixture $\sigma$.

Eq.(\ref{appen-5}) can be used to compute the upper bound of the three-tangle for the higher-rank
states. For example, let us consider the following rank-$8$ state
\begin{equation}
\label{rank-8-1}
   \rho
      =
         \xi \sigma + (1 - \xi) \tilde{\sigma}
\end{equation}
where $\sigma$ is given in Eq.(\ref{rank-4-1}) and $\tilde{\sigma}$ is 
\begin{equation}
\label{rank-8-2}
   \begin{array}{l}
      \tilde{\sigma}
         =
            y
            |GHZ,1-\rangle \langle GHZ,1-| 
                                                  \\
          \qquad
          + \frac{1-y}{3}
            \big[
               \;
                  |GHZ,2-\rangle \langle GHZ,2-|
                                                  \\
                \qquad\qquad\quad
                + |GHZ,3-\rangle \langle GHZ,3-|
                                                  \\
                \qquad\qquad\quad
                + |GHZ,4-\rangle \langle GHZ,4-| 
               \;
            \big].
   \end{array}
\end{equation}
If $x=y$, $\sigma$ and $\tilde{\sigma}$ are local-unitary(LU) equivalent with each other. Since
the three-tangle is LU-invariant quantity, $\tau_3(\tilde{\sigma})$ should be identical to 
$\tau_3 (\sigma)$ when $x=y$

Since $\rho$ is rank-$8$ mixed state, it seems to be extremely difficult 
to compute its three-tangle
analytically. If, however, $0 \leq y \leq 3/4$, $\tau_3(\tilde{\sigma})$ becomes zero and 
the above analysis yields a non-trivial upper bound of $\tau_3(\rho)$ as following:
\begin{equation}
\label{rank-8-3}
\tau_3(\rho) \leq \xi \tau_3 (\sigma).
\end{equation}

In this letter we have shown that the three-tangle does not properly quantify the three-party
entanglement for some mixture composed of only GHZ states. This fact has been used to
compute the (upper bound of) three-tangles for the higher-rank mixed states.

The fact $\tau_3 (\sigma) = 0$ for $x \leq 3/4$ can be used to find other 
rank-$4$ mixtures which have vanishing three-tangle by considering the Bloch 
hypersphere of $d=4$ qudit system.
First, we correspond the GHZ-states in $\sigma$ to the basis of the qudit system as 
follows:
\begin{equation}
\label{boso-3}
   \begin{array}{l}
      |GHZ,1+\rangle = \left( 1,0,0,0 \right)^T, \;
      |GHZ,2+\rangle = \left( 0,1,0,0 \right)^T
                                                   \\
      |GHZ,3+\rangle = \left( 0,0,1,0 \right)^T, \;
      |GHZ,4+\rangle = \left( 0,0,0,1 \right)^T
   \end{array}
\end{equation}
where $T$ stands for transposition.
It is well-known\cite{ber08} that the density matrix of the arbitrary $d=4$ qudit state can be 
represented by $\rho = (1/4) (I + \sqrt{6} \vec{n} \cdot \vec{\lambda})$, where $\vec{n}$ is 
$15$-dimensional unit vector and 
\begin{equation}
\label{boso-4}
   \vec{\lambda}
      = 
         \left(
               \Lambda_s^{12}, \cdots, \Lambda_s^{34},
               \Lambda_a^{12}, \cdots, \Lambda_a^{34}, 
               \Lambda^1, \Lambda^2, \Lambda^3
         \right).
\end{equation}
The generalized Gell-Mann matrices $\Lambda_s^{ij}$, $\Lambda_a^{ij}$ and $\Lambda^j$ are 
explicitly given in Ref.\cite{ber08}.
Then, the $15$-dimensional Bloch vectors for $|X \left(3/4, 0, 0, 0\right) \rangle$, 
$|X \left(3/4, 0, \pi, \pi \right) \rangle$,
$|X \left(3/4, \pi, 0, \pi \right) \rangle$, and 
$|X \left(3/4, \pi, \pi, 0\right) \rangle$ can be easily derived.
Thus, these four points form a hyper-polyhedron in $16$-dimensional space. Then all rank-4 quantum
states corresponding to the points in this hyper-polyhedron have vanishing three-tangle.

As we have shown in this letter, $\Pi_{GHZ}$ has vanishing two and three-tangle, but non-vanishing
one-tangle. It makes the left-hand side of the monogamy inequality 
$\tau_{3} + {\cal C}^2_{AB} + {\cal C}^2_{AC} \leq {\cal C}^2_{A(BC)}$ reduce zero. Thus, 
natural question arises: what physical resources make the one-tangle to be non-vanishing? 
Authors in Ref.\cite{bai07-1} conjectured that the origin of the non-vanishing one-tangle
comes from the higher tangles of the purified state. To support their argument they considered
a multipartite entanglement measure defined
\begin{equation}
\label{QCR}
E_{ms} (\Psi_N) = \frac{\sum_k \tau_{k(R_k)} - 2 \sum_{i < j} {\cal C}_{ij}^2}{N}
\end{equation}
where $\tau_{k(R_k)} = 2 (1 - \mbox{Tr} \rho_k^2)$ and $|\Psi_N\rangle$ is a $N$-qubit
purified state of the given mixed state. Since the numerator of $E_{ms}$ is difference between 
the total one-tangle and total two-tangle, it measures a contribution of the higher-tangles
to the one-tangle.
If we choose the purified state as 
\begin{equation}
   \begin{array}{l}
   |\Psi_5\rangle
      =
         \frac{1}{\sqrt{3}}
         |GHZ,2+\rangle |00\rangle
       + \frac{1}{\sqrt{3}}
         |GHZ,3+\rangle |01\rangle
                                        \\
    \phantom{|\Psi_5\rangle =}
       + \frac{1}{\sqrt{3}}
         |GHZ,4+\rangle |10\rangle, 
   \end{array}
\end{equation}
$E_{ms} (\Psi_5)$ reduces to $43/45$, which is larger than the one-tangle $5/9$. Thus,
it is possible that part of $E_{ms} (\Psi_5)$ converts into the non-vanishing one-tangle. 
However, still we do not know how to compute the one-tangle explicitly from $E_{ms} (\Psi_5)$.

The three-tangle itself is a good three-party entanglement measure. It exactly coincides with
the modulus of a Cayley's hyperdeterminant\cite{cay1845} and is polynomial 
invariant under the local $SL(2,\mathbb{C})$ transformation\cite{ver03}. As shown, however, 
it cannot properly quantify the three-party entanglement of W-state and $\Pi_{GHZ}$: 
$\tau_3 (W) = \tau_3 (\Pi_{GHZ}) = 0$. On the other hand, the $\pi$-tangle gives the non-zero
values: $\pi_3 (W) = 4 (\sqrt{5} - 1) / 9$ and $\pi_3 (\Pi_{GHZ}) = 1/9$. Does this fact simply
imply the crucial defects of the three-tangle as a three-party entanglement measure? 
Here, we would like to comment on the physical implication of $\tau_3 (\Pi_{GHZ}) = 0$. Few 
years ago the three-qubit mixed states were classified in Ref.\cite{acin01}. Following 
Ref.\cite{acin01} the whole mixed states are classified as separable (S), biseparable (B),
W and GHZ classes. These classes satisfy $S \subset B \subset W \subset GHZ$. 
One remarkable fact, which was proved in this reference, is that the
$W \setminus B$ class is not of measure zero among
all mixed-states. This is contrary to the case of the pure states, where the set of W-state 
forms measure zero\cite{dur00-1}. This fact implies that 
the portion of $W \setminus B$ class in the 
whole mixed states becomes larger compared to that of W class 
in the whole pure states. How could 
this happen? The fact $\tau_3 (\Pi_{GHZ}) = 0$ sheds light on this issue. Since 
$\Pi_{GHZ}$ has zero three-tangle but non-zero $\pi$-tangle, it is manifestly an 
element of $W \setminus B$ class.
As shown in Eq.(\ref{tangle-x-1}), however, it consists of three 
GHZ states without pure W-type state. We think there are many $W \setminus B$ states, 
which are mixture
of only GHZ states. It increases the portion of $W \setminus B$ class and eventually makes the 
$W \setminus B$ class to be of non-zero measure in the whole mixed states. 

{\bf Acknowledgement}: 
This work was supported by the Kyungnam University
Foundation Grant, 2008.

\end{document}